\documentclass[amsmath,amssymb,aps,prl,groupedaddress,floatfix,12pt]{revtex4-1}

\usepackage{graphicx}
\usepackage{color}
\usepackage{natbib}
\usepackage{amssymb}
\usepackage{amsmath}
\usepackage{amsfonts}
\usepackage{mathrsfs}
\usepackage{xcolor}
\usepackage{enumerate}
\usepackage{lineno}

\begin{document}


\title{Unsighted deconvolution ghost imaging}

\author{Yuan Yuan}
\author{Hui Chen}
\email[]{Corresponding author. chenhui@xjtu.edu.cn}
\thanks{These two authors contributed equally.}
\affiliation{Electronic Material Research Laboratory, Key Laboratory of the Ministry of Education and International Centre for Dielectric Research, Xi'an Jiaotong University, China,710049}
\affiliation{These two authors contributed equally.}


\date{\today}

\begin{abstract}
Ghost imaging (GI) is an unconventional imaging method that retrieves the image of an object by correlating a series of known illumination patterns with the total reflected (or transmitted) intensity. We here demonstrate a scheme which can remove the basic requirement of knowing the incident patterns on the object, enabling GI to non-invasively image objects through turbid media. As an experimental proof, we project a set of patterns towards an object hidden inside turbid media that scramble the illumination, making the patterns falling on the object completely unknown. We theoretically prove that the spatial frequency of the object is preserved in the measurement of GI, even though the spatial information of both the object and the illumination is lost. The image is then reconstructed with phase retrieval algorithms. 
\end{abstract}


\maketitle

\section{Introduction}
Different than the conventional imaging methods that rely on the first-order interference (typically using lenses), Ghost imaging(GI) exploits the second-order correlation to reconstruct an image\cite{Pittman1995}, bringing advantages such as better resistance to turbulence\cite{yang_lensless_2016}, high detection sensitivity\cite{morris2015imaging,UltraHighSpeedGI}, lensless imaging capability\cite{scarcelli2006can}, and broad adaptability for different scenarios\cite{Peng2015The,Liu2017Computational,Aspden2015Photon}. Therefore, GI has drawn a lot of attention during the past two decades, invoking a lot potential applications in various fields ranging from optical imaging\cite{gong2016three}, X-ray imaging\cite{pelliccia2016experimental,yu2016fourier,zhang2018tabletop}, to atomic sensing\cite{khakimov2016ghost,baldwin2017ghost}. 

A typical GI setup consists of a test and a reference arms. In the test arm, an object is illuminated by a temporally and spatially changing light field. The reflected or transmitted light from the object is collected by a bucket detector with no spatial-resolving capability. The reference arm is used to measure the variance of the light field on a conjugate plane of the object plane, i.e., the illumination patterns. The correlation between the patterns and the bucket signal results in the image of the object. Computational ghost imaging (CGI) is a setup that removes the reference arm. Instead, it pre-calculates or pre-determines the light patterns on the object plane. Since the bucket detector is simply used for collecting all the light from the object, the test arm is resistant to turbulence or strong light scattering. Most works focused on investigating GI with a turbid medium placed between an object and a bucket detector\cite{tajahuerce_image_2014,yang_lensless_2016,bai_imaging_2017,ye_embedding_2020}. GI requires the light patterns illuminating the object  be well determined. If the illumination patterns are disturbed by weak scattering, the reconstruction quality will be degraded\cite{bina_backscattering_2013,li_compressive_2020}. If a strong-scattering medium placed between the source and the object totally scrambles the patterns, GI will fail to recover an image\cite{gong_correlated_2011,yang_lensless_2016}. Although there was an approach that used ghost imaging configuration to image objects hidden behind turbid media, it exploited the non-Gaussion correlation which results in the resolution of the same order as the distance from the front surface of a turbid medium to the object (the width of the medium plus the distance between object to the medium)\cite{paniagua-diaz_blind_2019}. If this distance is several centimeters, it could not resolve structures smaller than a centimeter, limiting its applications. This is different from a typical GI scheme which is based on Gaussian statistics and whose resolution depends on the grain size of the illuminating patterns on the object\cite{ZhuGI2005,fayard_intensity_2015}. 

We here propose a new scheme that allows one to perform ghost imaging without knowing the illumination patterns on the object. The imaging resolution still obeys the same rule in GI: it is determined by the average grain size of the illumination patterns. Our method enables GI to non-invasively image an object completely hidden inside opaque media. 
In the following, We demonstrate a GI experiment with an opaque medium presented between the light source and an object.  The image is deconvolved from the correlation between the preset patterns of the source and the bucket detections, after applying a phase retrieval algorithm. A theoretical proof is provided in the following section.

\section{Experiment}

\begin{figure}[hbt]
    \centering
    \includegraphics[width=120mm]{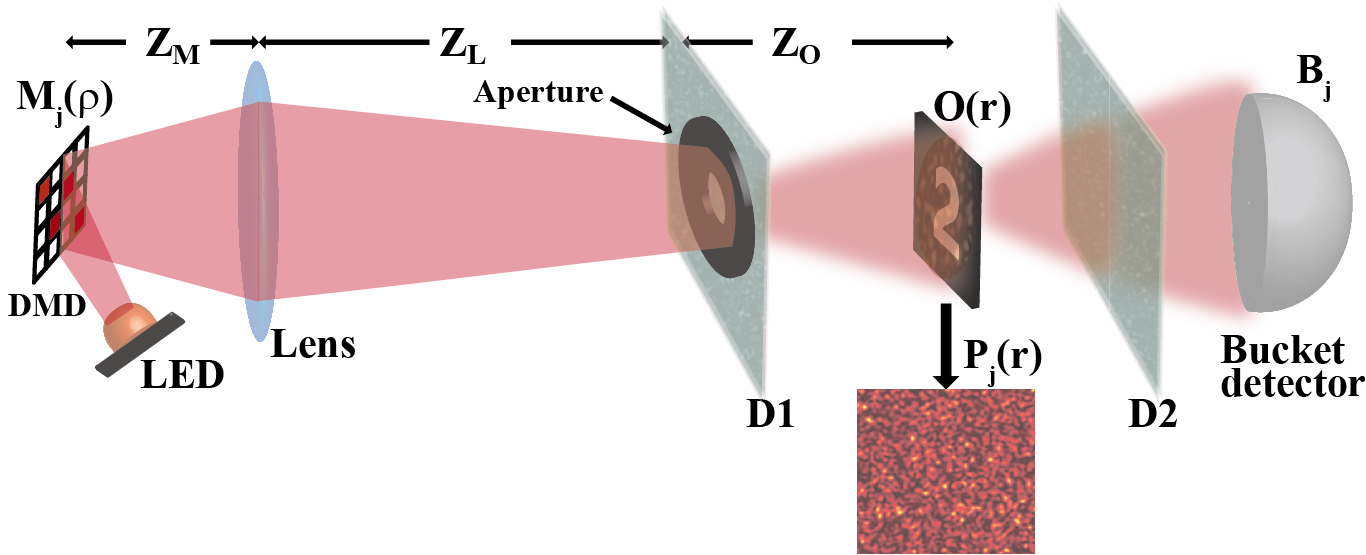}    
    \caption{\label{fig:SetupTTM} {Schematic of ghost imaging for an object hidden between two opaque diffusers. 
    A DMD (Digital Micromirror Device) displays a sequence of patterns $\{M_j({\boldsymbol \rho})\}$. A LED bulb and a lens are used to project the patterns towards an object hidden behind a diffuser (D1). D1 scatters the illuminating light and generates random speckles $\{P_j({\boldsymbol r})\}$ on object $O({\boldsymbol r})$. A small aperture with a diameter of $D\simeq 6\;mm$ is placed right in front of D1. A bucket detector measures the transmitted light from the object passing through another diffuser (D2), giving the bucket signals of $\{B_j\}$. The size of each cell is of the DMD $\sim 7.4\;\mu m$. The object is a hollow letter ``2'' of $\sim 0.8\times 0.5\;mm$. The focus length of the lens is $25\;mm$. $Z_M=70\;mm$. $Z_L=250\;mm$. $Z_O=300\;mm$. 
    } }    
\end{figure}

\begin{figure}[hbt]
    \centering
    \includegraphics[width=120mm]{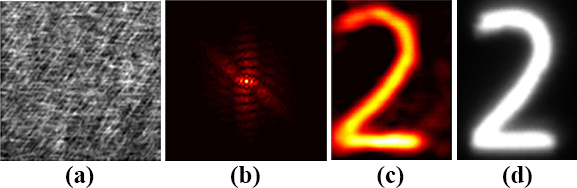}
\caption{\label{fig:ExpResult} {(a) The correlation between the preset patterns and the corresponding bucket signals, $C({\boldsymbol r})$; (b) the Fourier magnitude of (a), $|\tilde{C}({\boldsymbol u})|$; (c) The image reconstructed from  $|\tilde{C}({\boldsymbol u})|$ after using the phase retrieval algorithm which is a combination of the hybrid input-output (HIO) and the error reduction (ER) algorithms\cite{Fienup1978Reconstruction};  (d) The direct image of the object. }}
\end{figure}

The experimental setup is sketched in Fig. \ref{fig:SetupTTM}.  We used $1024\times 1024$ cells of the DMD to project $\sim 10^6$ preset patterns, $\{M_j({\boldsymbol \rho})\}$, towards the object. The diffuser D1 placed between the light source and the object scatters the projecting patterns and forms random speckle-like illuminating patterns on the object plane which is denoted as $\{P_j({\boldsymbol r})\}$. As illustrated in the inset of Fig. \ref{fig:SetupTTM}, $\{P_j({\boldsymbol r})\}$ become completely indeterminable due to strong scattering. The corresponding bucket signals were successively recorded.  The correlation between $\{M_j({\boldsymbol \rho})\}$ and the corresponding bucket signals, $C({\boldsymbol r})$, yields a random pattern (see Fig. \ref{fig:ExpResult}(a)) rather than the image of the object as in a typical GI system.  However, as long as the size of the light source is small enough (satisfying the isoplanatic condition for the scattering layers), the Fourier magnitude of $C({\boldsymbol r})$,  denoted as $|\tilde{C}({\boldsymbol u})|$ (see Fig. \ref{fig:ExpResult}(b)), exhibits the spatial frequency of the object. Applying a phase retrieval algorithm, we eventually obtained the image of the object (Fig. \ref{fig:ExpResult}(c)).  A theoretical interpretation is given in the following section.

\section{Theory}
In the following, we provide a general theoretical analysis for ghost imaging system with incoherent source in terms of PSFs. If the size of the light source satisfies the isoplanatic condition, i.e., it is within the memory effect range with respective to the scattering medium (D1 in Fig. \ref{fig:SetupTTM}), the PSF from the DMD plane to the object plane, $S_{MO}({\boldsymbol r}-{\boldsymbol \rho})$,  is shift-invariant\cite{Freund1988Memory}. The resulting illumination pattern on the object plane can be described by the linear system theory, which is written as the convolution of the source and the PSF, i.e., $P_j({\boldsymbol r})=[M_j {\ast} S_{MO}]({\boldsymbol r})$. 
The bucket detection can be formulated as:
\begin{align}\label{eq:Bucket}    
    B_j &=\int O({\boldsymbol r}) \cdot P_j({\boldsymbol r}) d{\boldsymbol r}=\int O({\boldsymbol r}) \cdot \int M_j({\boldsymbol \rho}) S_{MO}({\boldsymbol r}-{\boldsymbol \rho})d{\boldsymbol \rho}d{\boldsymbol r}\cr
    &=\int \left[O {\ast} S_{MO}\right]({\boldsymbol \rho})\cdot M_j({\boldsymbol \rho})d{\boldsymbol \rho}.
\end{align}
 
Since the spatial information of the illumination on the object is lost due to scattering, the correlation between $\{P_j({\boldsymbol r})\}$ and the corresponding bucket signals cannot be established. Instead, we analyze the correlation between the preset patterns on the DMD and the bucket signals:
\begin{align}\label{eq:CM}    
    C({\boldsymbol \rho'})&\equiv \sum_j B_j \cdot M_j({\boldsymbol \rho'})=\int \left[O {\ast} S_{MO}\right]({\boldsymbol \rho})\sum_j M_j({\boldsymbol \rho'})M_j({\boldsymbol \rho})d{\boldsymbol \rho}\cr
    &=\int \left[O {\ast} S_{MO}\right]({\boldsymbol \rho})\cdot \delta_D({\boldsymbol \rho'}-{\boldsymbol \rho})d{\boldsymbol \rho}=\big[\left[O {\ast} S_{MO}\right]{\ast}\delta_D\big]({\boldsymbol \rho'}).
\end{align}
Here, $\delta_D$ is a peak function with a width that represents the resolution of $\{M_j({\boldsymbol \rho})\}$ (i.e., the pixel size of the DMD). Taking the Fourier transform of both sides, we obtain
\begin{align}\label{eq:FTCM}    
    \tilde{C}({\boldsymbol u})=\tilde{O}({\boldsymbol u})\cdot \tilde{S}_{MO}({\boldsymbol u})\cdot \tilde{\delta}_{D}({\boldsymbol u}),
\end{align} 
where the tilde denotes the two-dimensional Fourier transform, ${\boldsymbol u}$ is the spatial-frequency coordinate vector. $\tilde{S}_{MO}$ is called the optical-transfer function (OTF). The Fourier magnitude form of Eq. (\ref{eq:FTCM}) is 
\begin{align}\label{eq:FMCM}    
    |\tilde{C}({\boldsymbol u})|=|\tilde{O}({\boldsymbol u})|\cdot |\tilde{S}_{MO}({\boldsymbol u})|\cdot |\tilde{\delta}_{D}({\boldsymbol u})|.
\end{align}
$|\tilde{S}_{MO}({\boldsymbol u})|$ is the magnitude of the OTF (MTF). Note that $\tilde{C}({\boldsymbol u})$ and $\tilde{\delta}_{D}({\boldsymbol u})$ can be easily determined. As long as $|\tilde{S}_{MO}({\boldsymbol u})|$ is predictable, the spatial frequency of the object, $|\tilde{O}({\boldsymbol u})|$, can be resolved. 

Eq. (\ref{eq:Bucket})-(\ref{eq:FMCM}) work for a general GI system with incoherent source. In the following, we discuss two cases without or with turbid media, respectively, in which the property of $|\tilde{S}_{MO}({\boldsymbol u})|$ can be predicted:

{\it Case 1}: {\bf without the turbid medium} (namely D1 is removed). Here, the patterns on the DMD can be precisely projected on the object with the lens. In a lens system, $\tilde{S}_{MO}$ is determined only by the pupil function of the lens, $\tilde{S}_{MO}\propto T_L^{1/2}{\ast}T_L^{1/2}$,  where $T_L$ is the squared modulus of the pupil function. From Eq.  (\ref{eq:FTCM}), $\tilde{C}({\boldsymbol u})$ is the Fourier transform of the object filtered by $\tilde{S}_{MO}$ and $\tilde{\delta}_{D}$. In another word, $C({\boldsymbol \rho'})$ is the diffraction-limited image of the GI system.
$|\tilde{S}_{MO}|$ and $|\tilde{\delta}_{D}|$ act as spatial frequency filters, defining the resolution of the image. 

{\it Case 2}: {\bf with strong-scattering medium}. Here, the projecting patterns are scrambled by the diffuser and become speckle-like patterns. Neither ${S}_{MO}$ can be determined, nor the image can be directly recovered from a correlation measurement. To solve this problem, we write the PSF into the convolution of two successive PSFs: $S_{MO}({\boldsymbol r}-{\boldsymbol \rho})=[S_{L}({\boldsymbol \xi}-{\boldsymbol \rho}){\ast} S_{S}({\boldsymbol r}-{\boldsymbol \xi})]({\boldsymbol r}-{\boldsymbol \rho})$. Here, ${\boldsymbol \xi}$ is the coordinate vector of an arbitrary transvers plane located in between the lens and the object. $S_L({\boldsymbol \xi}-{\boldsymbol \rho})$ is the PSF of the lens system from the DMD plane to ${\boldsymbol \xi}$'s plane. $ S_{S}({\boldsymbol r}-{\boldsymbol \xi})$ is the PSF of the scattering system from the ${\boldsymbol \xi}$'s plane to the object plane.  The Fourier magnitude form is, 
\begin{align}\label{eq:FTCMSCT}    
    |\tilde{C}({\boldsymbol u})|=|\tilde{O}({\boldsymbol u})|\cdot |\tilde{S}_{L}({\boldsymbol u})|\cdot |\tilde{S}_{S}({\boldsymbol u})| \cdot |\tilde{\delta}_{D}({\boldsymbol u})|.
\end{align} 
$|\tilde{S}_S|$  plays the same role as $|\tilde{S}_L|$ does, i.e., limits the range of the spatial frequency. The product of the three MTFs, $F({\boldsymbol u})\equiv |\tilde{S}_{L}({\boldsymbol u})|\cdot |\tilde{S}_{S}({\boldsymbol u})|\cdot|\tilde{\delta}_{D}({\boldsymbol u})|$, acts as an overall spatial filter which cuts off the high spatial frequency, defining the resolution of the image. Based on the above analysis we can see that, $|\tilde{C}({\boldsymbol u})|=|\tilde{O}({\boldsymbol u})|\cdot F({\boldsymbol u})$ is the Fourier magnitude of a diffraction-limited image.

When the scattering system contains sufficient random scatterers to satisfy the ergodic-like condition\cite{Freund1990Looking,Yuan_2020}, 
\begin{equation}\label{eq:PSF_FD}
      |\tilde{S}_S({\boldsymbol{u}})|\propto \left\{ [T_S\star T_S]({\boldsymbol{u}})+ \delta_G \right\}^{1/2},
\end{equation} 
where $\delta_G$ is a delta-like peak, representing the zero frequency relating to the background of the illumination. $T_S$ represents the squared modulus of the aperture function right in front of the D1. With an even light intensity attenuation over the circular-shape diffuser (as in our experiment), $|\tilde{S}_S({\boldsymbol{u}})|$ has a gentle slope within the  spatial-frequency range of [-$\frac{D}{\lambda Z_{O}}$, $\frac{D}{\lambda Z_{O}}$]\cite{Yuan_2020}. 
$|\tilde{S}_S|$ works as a spatial frequency filter as $|\tilde{S}_{L}|$ and $|\tilde{\delta}_{D}|$ do. In our experiment, the resolution of the image is mainly limited by  $|\tilde{S}_S|$, i.e., $\sim \lambda Z_{O}/D$.  
The missing Fourier phase of the image can be restored from $|\tilde{C}({\boldsymbol u})|$ using a phase retrieval algorithm, such as HIO or ER\cite{Fienup1978Reconstruction}. The image is then reconstructed.

\section{Discussion}
We have demonstrated a GI experiment without knowing the light patterns illuminating on an object. Even if the object is hidden entirely inside strong-scattering media, the image can still be recovered as long as the light source is small enough. To interpret this imaging scheme, we provide a theoretical analysis for a general ghost imaging system with incoherent light in terms of PSFs. In the presence of a turbid medium between the source and the object, the correlation measurement of GI resolves the spatial frequency of the object (instead of an image of the object), from which a diffraction-limited image can be deconvolved with a phase retrieval algorithm, without knowing  the spatial information of the PSF and the patterns falling on the object. 

In this imaging scheme, the source should be within the memory effect range with respect to the turbid medium (D1). Otherwise, the speckles generated by scattering would be flattened out. It is not hard to fulfill this condition though: one can either physically shrink the size of the source, or employ lenses to produce a smaller virtual source, or move the source further away. This feature makes ghost imaging a candidate for non-invasive imaging through turbid media. 



\bibliography{MM5}

\end{document}